\documentclass[%
 aip,
 amsmath,amssymb,
 reprint,%
]{revtex4-1}

\usepackage{graphicx}
\usepackage{dcolumn}
\usepackage{bm}
\usepackage{multirow}
\usepackage{color}
\usepackage{amsmath}

\usepackage{textcomp}

\usepackage[utf8]{inputenc}
\usepackage[T1]{fontenc}
\usepackage{mathptmx}
\usepackage{etoolbox}

\usepackage[utf8]{inputenc}
\usepackage{mathptmx}
\usepackage{etoolbox}

\newcommand{\edit}[1]{{\textcolor{black}{{#1}}}}

\makeatother
\begin{document}

\preprint{AIP/123-QED}

\title[]{Lattice Boltzmann models for the hydrodynamic equations in multiphase flow with high density ratio}

\author{H. Otomo }
\thanks{Author to whom correspondence should be addressed: Hiroshi.OTOMO@3ds.com}
%
\author{C. Sun }
%
\author{T. Inamuro }
\altaffiliation[Also at ]{Graduate School of Engineering, Kyoto University, Kyoto 615-8540, Japan}
\author{W. Li}
%
\author{M. Dressler}
%
\author{H. Chen}
\altaffiliation[Also at ]{Department of Mechanical Engineering and Mechanics, Lehigh University, Pennsylvania 18016, USA}
\author{Y. Li }
%
\author{R. Zhang} 
\affiliation{ Simulia R\&D, Dassault Systémes Americas Corp., Waltham, MA 02451,USA.}

\date{\today}

\begin{abstract}
Multiphase flows with high density ratios, such as water and air flows, have recently been simulated using the lattice Boltzmann (LB) method. This approach corresponds to solving the phase field equations, such as the Cahn-Hilliard and Allen-Cahn equations, and the hydrodynamic equations, typically the Navier-Stokes and pressure equations for pseudo-incompressible fluids.
Due to the high density ratio, the higher-order numerical truncation errors associated with spatial density gradients can become significant.
These errors can lead to problems such as inaccuracies in shear stress, violations of Galilean invariance, and undesirable dependencies on absolute pressure for the pseudo-incompressible solutions.
To overcome such problems, the moments of the distribution function and the equilibrium state must be carefully designed while ensuring robustness.
In this work, we propose a new scheme based on the lattice kinetic scheme (LKS), which directly solves the velocity and pressure fields in the similar discrete space as the LB method. 
When mapping the LKS-based models to the LB models, the original LKS models are simplified for computational efficiency and the filter collision operator is implemented.
Benchmark test cases confirm that the proposed scheme effectively addresses these issues, achieving high accuracy and robustness while eliminating the iterative steps typically required in the LKS. 
One of the most significant improvements is the accuracy of the airflow field induced by water motion, likely due to improved momentum transfer across the interface.
\end{abstract}

\maketitle

{\bfseries Copyright 2025 American Institute of Physics. This article is accepted manuscript version and is not the final published version.}


\section{Introduction}
\label{Intro}
After decades of dedicated research on the lattice Boltzmann (LB) method, the simulation of multiphase flows with high density ratios is becoming feasible with reasonable robustness and accuracy.
One promising approach is to solve two coupled LB equations; one for the phase field whose dynamics is governed by the Cahn-Hilliard equation, the conservative Allen-Cahn equation, or its extended version.\cite{sun2007sharp,chiu2011conservative,JAIN2022111529, Wang_review_2019}
The other is for hydrodynamics, characterized by the pressure field $P$ and the velocity field $u_{\alpha}$. 
In the case of pseudo-incompressible fluids, the governing macroscopic equations, \edit{the pressure equation and the Navier-Stokes equation,} can be written as
\begin{align}
\label{prs_eq}
\frac{\partial P}{\partial t} + \rho T \frac{\partial u_{\beta}}{\partial x_{\beta}}=0,  \\
\frac{\partial u_{\alpha}}{\partial t} + \frac{\partial u_{\alpha} u_{\beta}}{\partial x_{\beta}} = -\frac{1}{\rho} \frac{\partial P}{\partial x_{\alpha}}  + \frac{1}{\rho} \frac{\partial \sigma_{\alpha, \beta}}{\partial x_{\beta}} + \frac{F_{\alpha}}{\rho}, \label{mom_eq}
\end{align}
where $\rho$ is the density, $T$ is the temperature, and $F_{\alpha}$ is the force field including the surface tension force.
The indices $\alpha, \beta \in \left\{ x, y ,z \right\}$ are for the Cartesian coordinates.
This paper assumes summation for all repeated indices. 
The stress tensor $\sigma_{\alpha, \beta}$ is,
\begin{align}
\sigma_{\alpha, \beta} = \rho \nu \left( \frac{\partial u_{\alpha}}{\partial x_{\beta}}  + \frac{\partial u_{\beta}} {\partial x_{\alpha}}  \right)
+ \rho \frac{2}{3} \left( \lambda_\mathrm{b} - \nu \right) \frac{\partial u_l}{\partial x_l} \delta_{\alpha, \beta},
\end{align}
with the kinematic viscosity $\nu$ and the bulk viscosity $\lambda_\mathrm{b}$.
It is easy to see that the combination of Eq.~(\ref{prs_eq}) and Eq.~(\ref{mom_eq}) leads to the sound wave equation with the sound speed of $\sqrt{T}$, which is consistent with the typical single-phase LB models. \cite{Sauro_book2001}
Therefore, the pressure field is typically relaxed with the time scale of $L/\sqrt{T}$ where $L$ is the characteristic length.
After this timescale, the velocity field approximately follows the solution of the incompressible fluids. 
Compared to the compressible fluid framework, such a pseudo-incompressible assumption is advantageous to deal with the artificial compressibility problem of the LB methods, which is usually observed in internal flows with increased simulation Mach number. \cite{otomo2024bdfrcmethodology} 
In this study, we discuss the LB models that lead to the hydrodynamic equations of Eq.~(\ref{prs_eq}) and Eq.~(\ref{mom_eq}).

The LB model is formulated in such a way that Eq.~(\ref{prs_eq}) and Eq.~(\ref{mom_eq}) are derived from the LB equations in the leading order using certain ordering methods, such as the Chapman-Enskog expansion or the Taylor expansion, \edit{for example.} \cite{OTOMO20171000}
For example, the LB equation with the BGK-collision operator,
\begin{align}
\label{Org_LB}
f_i \left( x_{\alpha} + c_{i, \alpha}  , t +1 \right) - f_i \left( x_{\alpha}   , t  \right) = - \frac{f_i - f^\mathrm{eq}_i}{\tau} \vert_{x_{\alpha}   , t },
\end{align}
is considered where $\tau$ is the relaxation time and $c_i $ is the discrete particle speed in the lattice direction $i \in$ [0,18] where the D3Q19 is employed. Here, $f_i$ is the distribution function and $\Delta t =1$ is assumed.
Redefining $\vec{x}$ and $t$ properly and substituting $f_i$ formula recursively, \cite{OTOMO20171000,OTOMO2019} we obtain 
\begin{align}
\label{feq_expansion}
f_{i} \left( x_{\alpha},t \right)
=
f^\mathrm{eq}_{i} \left( x_{\alpha} -c_{i, \alpha}, t-1 \right) \nonumber \\
+ \sum_{n=1}^{\infty} \left( 1 - \frac{1}{\tau} \right)^n   
\left\{ f^\mathrm{eq}_{i} \left( x_{\alpha}- \left( n+1 \right) c_{i, \alpha}, t - \left( n+ 1\right) \right)  \right. \\ \nonumber
\left.   - f^\mathrm{eq}_{i} \left( x_{\alpha} -n c_{i, \alpha}, t-n \right) \right\},
\end{align}
\edit{where $f_i$ in the initial condition is assumed to be the equilibrium state}.
Where $\tau=1$, using the Taylor expansion, Eq.~(\ref{feq_expansion}) can be written as,
\begin{align}
\label{feq_taylor expansion}
f_{i} \left( x_{\alpha},t \right)
=
\sum_{l=0}^{l=\infty} \frac{ \left( -1\right)^l}{l !} \left( \frac{\partial}{\partial t} + c_{i, \alpha}  \frac{\partial}{\partial x_{\alpha}} \right)^l f^\mathrm{eq}_i,
\end{align}
with the assumption that $f^\mathrm{eq}_i$ is analytic.
Taking the zeroth and first moments of Eq.~(\ref{feq_taylor expansion}) should lead to Eq.~(\ref{prs_eq}) and Eq.~(\ref{mom_eq}) in the leading order.
Through this algebra, we see two observations;
\begin{itemize}
\item All derivative terms can only be written with $f^\mathrm{eq}_i$, which is usually formulated with the macroscopic quantities.
\item The truncation error terms in Eq.~(\ref{prs_eq}) and Eq.~(\ref{mom_eq}) are formed by the derivatives of the  moments of $f^\mathrm{eq}_i$.
\end{itemize}
 These observations are likely to remain unchanged even when using $\tau \neq 1$ and/or the other collision operators, such as the filter collision operator, as long as the recursive substitution of $f_i$ can be applied with the reasonable ordering.
 
\begin{figure}
\includegraphics[width=0.9\linewidth]{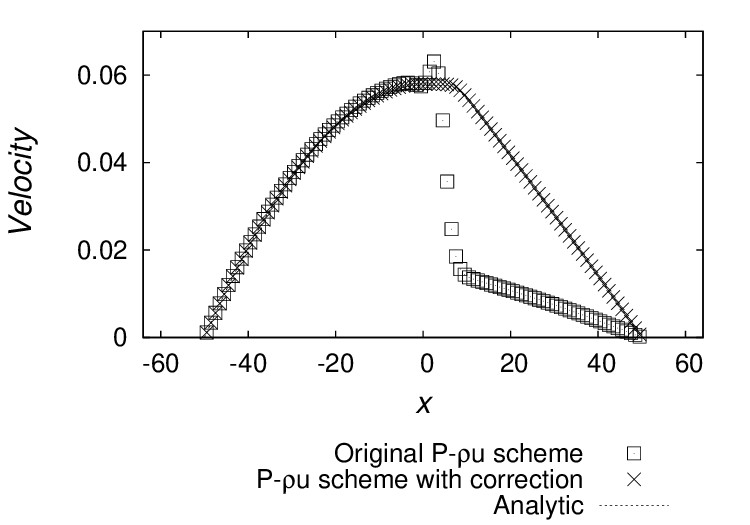}
\caption{ Velocity profiles of the two-layer Poiseuille flow using the original $P \mbox{-} \rho u$ scheme  \cite{fakhari2016mass} and the $P \mbox{-} \rho u$ scheme with higher order corrections \cite{otomo2019improved}  are plotted with the square and cross symbols, respectively. The analytical solution is represented by a line. }
\label{fig_poiseuille_P_rhou}
\end{figure}

\begin{figure}
\includegraphics[width=0.9\linewidth]{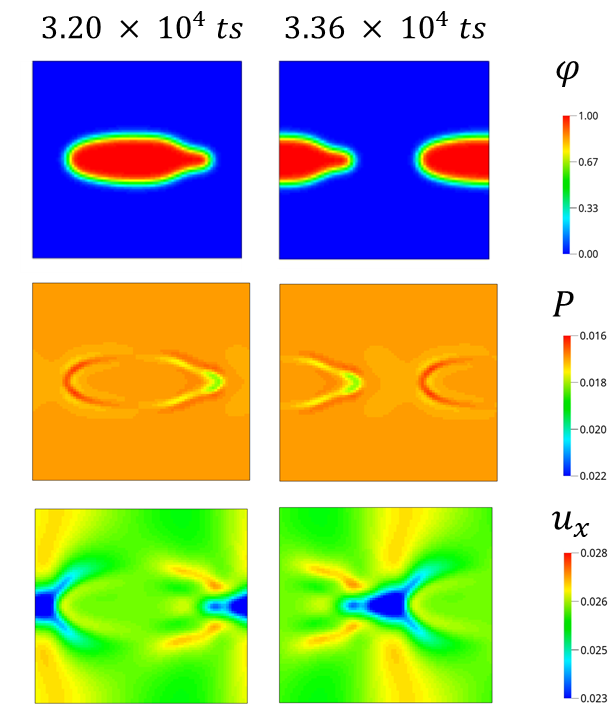}
\caption{ The order parameter $\varphi$ profiles at 3.20$\times 10^4$  (left) and 3.36$\times 10^4$ (right) timesteps using the  $P \mbox{-} \rho u$ scheme \cite{fakhari2017improved} for a case where velocity of 0.025 in the $x$-direction is initially applied everywhere. Similarly, the pressure and velocity in the $x$-direction are shown at the center and bottom. }
\label{fig_dynamicdro}
\end{figure}

\begin{figure}
\includegraphics[width=0.9\linewidth]{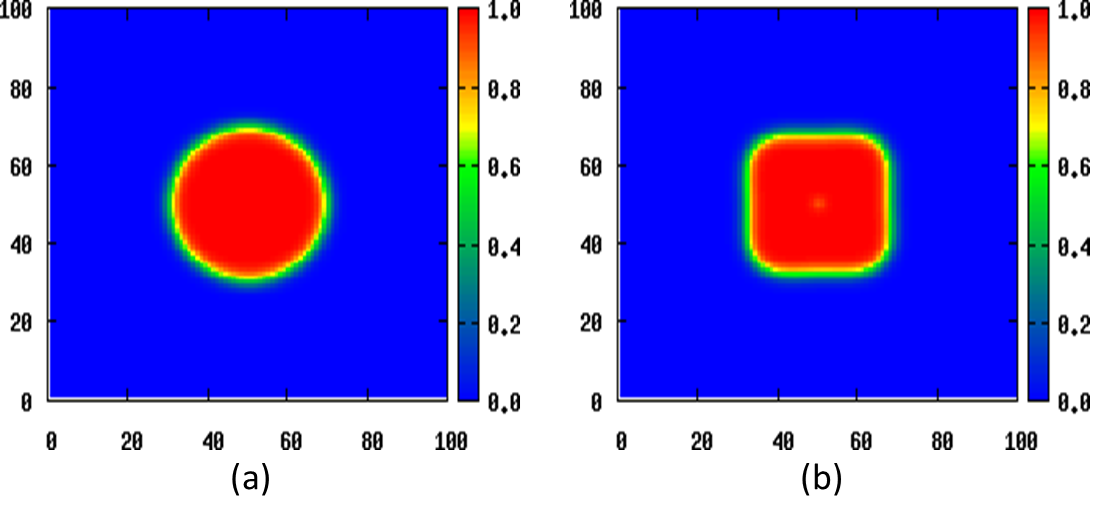}
\caption{ The order parameter $\varphi$ profiles using the  $P/\rho \mbox{-} u$ model \cite{fakhari2017improved} where the initial pressure is set to 0.01 (left) or 1.0(right) everywhere. }
\label{fig_static_dro_oldv}
\end{figure}

Let us consider two typical hydrodynamic LB models.
In the previous studies, \cite{Liang_2014,fakhari2016mass, fakhari2017diffuse, otomo2019improved} the LB model is formulated  in such a way that the zeroth and the first moment of $f^\mathrm{eq}_i$ lead to $P$ and $\rho u_{\alpha}$ respectively;
\begin{align}
\sum_i f^\mathrm{eq}_i = \frac{P}{T} + \cdots,  \: \: \: \:
\sum_i f^\mathrm{eq}_i c_{i, \alpha}=\rho u_{\alpha} + \cdots.
\end{align}
%
In the other study, \cite{fakhari2017improved} the LB model is formulated in such a way that the zeroth and the first moment of $f^\mathrm{eq}_i$ lead to $P/\rho$ and $ u_{\alpha}$ respectively;
\begin{align}
\sum_i f^\mathrm{eq}_i = \frac{P}{\rho T} + \cdots,  \: \: \: \:
\sum_i f^\mathrm{eq}_i c_{i, \alpha}= u_{\alpha} + \cdots.
\end{align}
\edit{For simplicity, the dimensional constant such as $\rho_{ref}$ is abbreviated in these equations.}
Let us call them the $P \mbox{-} \rho u$ scheme and $P/\rho \mbox{-} u$ scheme in this paper.
Although Eq.~(\ref{prs_eq}) and Eq.~(\ref{mom_eq}) are derived in the leading order using both schemes, the truncation error terms are largely different.
In particular, in the $P \mbox{-} \rho u$ scheme, the truncation errors in Eq.~(\ref{mom_eq}) may include the high-order derivative terms of $\rho u$ such as
\begin{align}
\label{truncation_error_rhou}
\frac{\partial^n \rho u_{\alpha}}{\partial x_{\beta_1} \partial x_{\beta_2} \cdots \partial x_{\beta_n}}
= u_{\alpha} \frac{\partial^n \rho}{\partial x_{\beta_1} \partial x_{\beta_2} \cdots \partial x_{\beta_n} }  \nonumber \\
+ \frac{\partial u_{\alpha}}{\partial x_{\beta_1}  } 
\frac{\partial^{n-1} \rho}{ \partial x_{\beta_2} \partial x_{\beta_3} \cdots \partial x_{\beta_n} } +
\cdots,
\end{align} 
 where the indices $\alpha$ and $\beta_l$ of $1 \le l \le n$ are properly contracted and $n$ is the even number of $n  \ge 4$. 
 %
In the case of high density ratio, these terms around the phase interface may not be trivial.
For example, the second term on the right can cause inaccuracies in the shear stress around the interface.
To overcome this problem, in a previous study, \cite{otomo2019improved} the higher order correction term was added to the equilibrium state.
In the Poiseuille flow of the two-layer multiphase flow, the improvements can be clearly seen as shown in Fig~\ref{fig_poiseuille_P_rhou} where the cross symbols show the improved velocity profile while the square symbol shows the original results compared to the analytical solution described by the solid line. 
Here in the two-dimensional channel of 100 lattice width, left half was occupied by water and the other half was occupied by air. 
The viscosities and surface tension are $\left\{ \nu_\mathrm{air}, \nu_\mathrm{water} \right\}= \left\{ 1.67 \times 10^{-1}, 1.10 \times 10^{-2} \right\}$ and $\sigma=1.0 \times 10^{-3}$. The density ratio and gravitational acceleration are $\rho_\mathrm{ratio}=1000$ and $g=5.0 \times 10^{-7}$.
Despite improvements with the corrections, it is not easy to remove all the problematic higher order terms. 
As another example, the first term on the right in Eq.~(\ref{truncation_error_rhou}) can break the Galilean invariance.
If a droplet whose radius is 16 is simulated in an $80 \times 80$ domain surrounded by the periodic boundaries with an initial constant velocity of 0.025 in the $x$-direction, the droplet changes its shape, and the velocity and pressure profiles deviate significantly from the constant profiles, as shown in the snapshots at two instances in Fig.~\ref{fig_dynamicdro}. 
This setup should indicate that the droplet is static in the moving frame, and therefore the results violate Galilean invariance. 
Here, the other settings such as viscosity and surface tension are the same as for the Poiseuille flow case. 
Similarly, with the $P/\rho \mbox{-}  u$ scheme, we may expect the truncation errors in Eq.~(\ref{prs_eq}) such as,
\begin{align}
\label{P-rho_truncation}
\frac{\partial^n P / \rho}{\partial x_{\beta_1} \partial x_{\beta_2} \cdots \partial x_{\beta_n}}
= P \frac{\partial^n 1/\rho}{\partial x_{\beta_1} \partial x_{\beta_2} \cdots \partial x_{\beta_n} } +\cdots,
\end{align} 
where the indices $\beta_l$ of $1 \le l \le n$ are properly contracted and $n$ is the even number of $n \ge 2$.
The first term on the right hand side can cause a dependence on the absolute pressure value, which is undesirable under the incompressible assumption.
As an example, the static droplet of radius 20 is simulated in the $100 \times 100$ domain surrounded by the periodic boundaries while initially setting the pressures everywhere to 0.01 or 1.0.
They show different converged droplet shapes as shown in Fig.~\ref{fig_static_dro_oldv}.
Here the viscosities are $3.6 \times 10^{-4}$ for both components,  $\sigma=2.5 \times 10^{-2}$, and $\rho_\mathrm{ratio}=1000$, respectively. 
The surface tension model is based on the chemical potential as shown in the latter section. 
This is one of the examples where the truncation error contaminates the surface tension effects depending on the absolute value of the ambient pressure. 

In this study, we formulate the LB model so that the zeroth and the first moment of $f^\mathrm{eq}_i$ lead to $P$ and $ u_{\alpha}$ respectively.
It is readily seen that such a $P\mbox{-}u$ scheme can avoid the problematic truncation errors discussed in Eq.~(\ref{truncation_error_rhou}) and Eq.~(\ref{P-rho_truncation}).
A possible obstacle to naively employing this scheme may be robustness.
On the other hand, in a previous study, \cite{INAMURO201655,INAMUROBOOK,INAMURO2021} the multiphase flow with high density ratio was simulated using the lattice kinetic scheme (LKS).
In the LKS, instead of solving the LB equation, the velocity and pressure fields are solved in discrete space whose connectivity between each adjacent fluid node is similar to the LB formulation.
Focusing on the correspondence, we convert their LKS-based models to the generic LB-based models while simplifying for the sake of computational efficiency. The conversion is helpful for incorporating many recent enhancements based on the LB-based models.
In fact, by implementing the filter collision operator, the robustness was improved and as a result we could skip the iteration steps \edit{introduced in the original LKS-based model while maintaining the sound speed of $\sqrt{T}$}.

This paper is organized as follows.
In Section~\ref{sec_formulation}, LB models for the multi-phase flow with high density ratio are introduced.
In Section~\ref{validation}, the LB models proposed in Section~\ref{sec_formulation} are validated against a set of benchmark test cases.
In Section~\ref{summary}, findings in this study are summarized.


\section{Lattice Boltzmann models} 
\label{sec_formulation}


Two lattice Boltzmann (LB) equations, one for the order parameter $\varphi$ and the other for the hydrodynamics, are typically solved for the simulation of multiphase flow with high density ratios.
In this study, we focus primarily on the discussion of hydrodynamic LB models.
For the LB model for $\varphi$, we use one corresponding to the conservative Allen-Cahn (AC) equation as an example. \cite{Geier_2015,fakhari2016mass,Ren_2016,Wang_2016}
According to a previous study, \cite{fakhari2016mass} it is formulated as,
\begin{equation}
\label{eq:LB_phi_filter_col}
 {h}_{i} \left( x_{\alpha} + c_{i, \alpha}, t+1 \right) =
 {h}_{i} \left( x_{\alpha}, t \right) - \frac{{h}_{i} - {h}^\mathrm{eq}_{i}}{\left(\frac{M}{T}\right) + 0.5} \Bigg|_{x_{\alpha} , t},
\end{equation}
where  ${h}^\mathrm{eq}_{i}$ is the equilibrium state defined as,
\begin{eqnarray}
{h}^\mathrm{eq}_{i} &=& \varphi \Gamma_i + \theta w_i \left( c_{i, l}  n_{l} \right), \\
\Gamma_i  &=& w_i \left\{ 1+ \frac{ c_{i, l}   u_{l} }{T} + \frac{\left(   c_{i, l}    u_{l}  \right)^2}{2 T^2} - \frac{ u^2}{2T} \right\}, \\
\theta &=& \frac{M}{T} \left\{ \frac{1- 4 \left( \varphi - 0.5 \right)^2 }{W} \right\}. \label{theta_phai_LBeq}
\end{eqnarray}
Here the index $l \in \left\{ x, y ,z \right\}$ and $i$ are for the Cartesian coordinates and discretized particle speeds, respectively. 
For example in the case of D3Q19, which is used in this paper, $i \in [ 1, 19]$ \edit{and $T=1/3$}. 
The summation of the distribution function $h_i$ for $i$ leads to the order parameter, namely $\sum_i {h}_i = \varphi$ and $\varphi$ basically ranges from 0 to 1.
The notation $M$, $W$, and $n_{\alpha}$ denotes the mobility, the interface thickness, and  the unit vector normal to the interface,  computed by $\nabla_{\alpha} \varphi / \left(  |\nabla \varphi|  + \epsilon \right)$ where $\epsilon$ is a small parameter taken as $10^{-10}$ in order to avoid division by zero. 
To avoid the rarefied droplets and dense bubbles, observed in dynamic cases with the conservative AC equation, $\theta$ in Eq.~(\ref{theta_phai_LBeq}) is corrected with the following $\delta \theta$; \cite{otomo2019improved, otomo2024methodology} 
\begin{align}
\delta \theta = \gamma \frac{M}{T W} \lvert \nabla \varphi \rvert \left\{ \left( 1+ \frac{\Delta \varphi}{\lvert \Delta \varphi + \epsilon \rvert} \right) Y_1 +\left( 1- \frac{\Delta \varphi}{\lvert \Delta \varphi + \epsilon \rvert} \right) Y_2    \right\}, \nonumber \\
Y_1= \min \left\{ \mathcal{F} \left( \frac{\varphi - \varphi_n}{D}\right), \mathcal{F} \left( - \frac{\varphi-1}{D} \right)\right\}, \nonumber \\
Y_2= \min \left\{ \mathcal{F} \left( \frac{\varphi }{D}\right), \mathcal{F} \left( - \frac{\varphi - \varphi_m}{D}\right) \right\}, \nonumber  \\
\mathcal{F}(x) = \max \left\{ 0, \min \left\{ 1, x\right\} \right\},
\end{align}
where $\gamma=-22.5$, $\varphi_n =0.8$, $\varphi_m=0.2$, $\epsilon = 1.0 \times  10^{-10}$, and $D=1.0 \times  10^{-4}$ for example. 
This correction adds the diffusivity in the location where $\Delta \varphi$ is sufficiently close to zero, $\lvert \nabla \varphi \rvert$ is non-zero, and $\varphi \in \left( 0, \varphi_m \right)$ or $\varphi \in \left( \varphi_n, 1 \right)$.
Note that such conditions are unlikely to be met for the regular interface regions, but are likely to be met for the rarefied droplets and dense bubbles. 
\edit{From these formulations, using certain analytical methods such as the Chapman-Enskog expansion, \cite{fakhari2016mass} one can find the corresponding macroscopic equation, 
\begin{align}
\label{AC_eq}
\frac{\partial \varphi}{\partial t} +  \frac{\partial \varphi u_{\beta}}{\partial x_{\beta}}= \frac{\partial}{\partial x_{\beta}} \left\{ M \left[ 1 -  \frac{\left( \theta +  \delta \theta \right)T}{M \left| \nabla \varphi \right|} \right] \frac{\partial \varphi}{\partial x_{\beta}} \right\}.
\end{align}
As discussed above, in the absence of artificially rarefied droplets and dense bubbles, $\delta \theta=0$.
In recent studies, \cite{Ren_2016,Wang_review_2019} the convection term is formulated as $u_{\beta} \partial \varphi / \partial x_{\beta}$ with correction terms at the expense of mass conservation or computational cost due to global integration.
In this study, the LB formulation for Eq.~(\ref{AC_eq}) is used under the assumption that the compressibility effect is negligible in such a convection term.
}

For the LB equation for the hydrodynamics, from the insight gained from the analysis in Section~\ref{Intro}, the LB models are formulated so that the zeroth and the first moment of $f^\mathrm{eq}_i$ and $f_i$ lead to $P$ and $ u_{\alpha}$ respectively  ($P \mbox{-} u$ scheme).\cite{otomo2024methodology_GI_Ap} 
To start from a promising model with robustness, we refer to the previous study based on the lattice kinetic scheme (LKS), \cite{INAMURO201655,INAMUROBOOK} in which the time evolution of velocity and pressure is calculated instead of solving the LB equation for the distribution function.  
If their formulation for velocity and pressure is regarded as the zeroth and first moments of the LB equation, it is possible to approximate their formulation to the LB formulation.
For smooth mapping and computational efficiency, we have replaced and removed some terms, such as those that can be replaced by the non-equilibrium states and require high-order derivative calculations. 
As a result, the LB equation for hydrodynamics can be written with the BGK collision operator as follows; 
\begin{align}
\label{BGK_LBeq_hydroeq}
g_{i} \left( x_{\alpha} + c_{i, \alpha}, t+1 \right) =
 g_{i} \left( x_{\alpha} , t  \right) - \frac{g_{i} - g^\mathrm{eq}_{i}}{\tau_\mathrm{mix}} \vert_{x_{\alpha}, t},
\end{align}
where
\begin{eqnarray}
g^\mathrm{eq}_{i}= \frac{P}{\rho_\mathrm{ref} T} \delta_{i, ist} + \Gamma_i - w_i + w_i \frac{c_{i, l} F_{l}}{\rho T } \nonumber \\
 + B \frac{w_i - \delta_{i, ist}}{T} \frac{\partial u_{l}}{\partial x_{l}} + w_i \frac{c_{i, l} \sigma_{l, m}}{\rho^2 T} \frac{\partial \rho}{\partial x_{m}}, \\
B=\frac{2}{3} \left( \nu - \lambda_\mathrm{b} \right) + \frac{T}{2}.
\end{eqnarray}
Here the index $ist$ is one such that $\left| c_{ist} \right|=0$ and $\rho_\mathrm{ref}$ is the characteristic density which is the light fluid's density $\rho_\mathrm{air}$ in this study.  
The force $F_{\alpha}$ includes the surface tension and the external force $F_\mathrm{ext}$ like the gravitational force,
\begin{eqnarray}
F_{\alpha} = F_{\mathrm{ext}, \alpha} + \mu \frac{\partial \varphi}{\partial x_{\alpha}},
\end{eqnarray}
where
\begin{eqnarray}
\mu=\frac{48 \sigma}{W} \varphi \left(  \varphi -1 \right) \left( \varphi -0.5 \right) - \frac{3 \sigma W}{2} \nabla^2 \varphi,
\end{eqnarray}
where $\sigma$ is the surface tension. The bulk viscosity $\lambda_\mathrm{b}$ is set to $1/6$.
The relaxation time $\tau_\mathrm{mix}$ is linearly interpolated using the air and water components of $\tau$ as,
\begin{equation}
\label{tau_interpolate}
\tau_\mathrm{mix} =\tau_\mathrm{air} + \varphi \left( \tau_\mathrm{water} -  \tau_\mathrm{air}  \right) ,
\end{equation}
where
\begin{eqnarray}
\tau_\mathrm{water} = \frac{\nu_\mathrm{water}}{T} + \frac{1}{2}, \: \: \: \tau_\mathrm{air} = \frac{\nu_\mathrm{air}}{T} + \frac{1}{2},
\end{eqnarray}
using the air and water kinematic viscosity,  $\nu_\mathrm{air}$ and $\nu_\mathrm{water}$.
For computational efficiency, the LB models are formulated so that all terms in Eq.~(\ref{BGK_LBeq_hydroeq}) require the derivative calculation up to the second order, which can be computed using the nearest neighbor cells with the central difference scheme,
\begin{eqnarray}
\frac{\partial \phi}{\partial x_{\alpha}} =\frac{ \sum_i  \left\{ \phi \left( x + c_{i} \right) -\phi \left( x - c_{i}  \right) \right\} c_{i, \alpha} w_i}{2T},  \\
\nabla^2 \phi = \frac{2 \sum_i \left\{ \phi \left(x + c_{i}  \right) -\phi \left(x \right) \right\} w_i}{T}, 
\end{eqnarray}
for a certain quantity $\phi$. 

The pressure and velocity are calculated as follows,
\begin{align}
P =\rho_\mathrm{ref} T \sum_{i} \left( g_i + \left( 1 + \omega \right) K_i  \right) + \omega \rho_\mathrm{ref} \sum_i w_i \left( c_{i, l} u_l \right) \vert_{x_{\alpha} - c_{i, \alpha} }, \\
u_{\alpha} = \sum_i \left( f_i + K_i \right) c_{i , \alpha},
\end{align}
where
\begin{eqnarray}
K_i = \frac{w_i}{T \rho} \left\{ P \left( x_{\alpha} - c_{i, \alpha} \right) - P \left( x_{\alpha} \right)  \right\}, \label{Kiterm} \\
\omega= \frac{\rho - \rho_\mathrm{air} }{\rho_\mathrm{water} - \rho_\mathrm{air}} \left( \omega_\mathrm{\max} - 1 \right).
\end{eqnarray}
Here, $\rho_\mathrm{water}$ is the characteristic density of the heavy fluid. 
In the original LKS model, \cite{INAMURO201655,INAMUROBOOK} a parameter $\omega_\mathrm{max}$ is set to $\rho_\mathrm{water}/ \left( n \rho_\mathrm{air} \right)$ where $n$ is the iteration number. 
In this study, since we do not have any iteration steps, $\omega_\mathrm{max}=\rho_\mathrm{water}/ \rho_\mathrm{air} = \rho_\mathrm{ratio}$. 
%

For the sake of robustness, Eq.~(\ref{BGK_LBeq_hydroeq}) is solved with the filtered collision operator, 
\begin{equation}
\label{eq:LB_filter_col}
 g_{i} \left( x_{\alpha}+c_{i, \alpha} , t+ 1 \right) =
 g_{i}^\mathrm{eq} \vert_{x_{\alpha}, t} + \left( 1 - \frac{1}{ \tau_\mathrm{mix} }  \right) \Phi_i : \Pi \vert_{x_{\alpha}, t},
\end{equation}
where $\Phi_i$ is a filtered operator that uses Hermite polynomials and $\Pi$ is the nonequilibrium moments of the momentum flux,
\begin{eqnarray}
\Phi_{i, \alpha \beta} = \frac{w_i}{2 T^2} \left( c_{i, \alpha} c_{i, \beta} - T \delta_{\alpha, \beta} \right), \label{momentum_flux0} \\
\Pi_{\alpha \beta} = \sum_{k}  c_{k, \alpha} c_{k, \beta} \left(   g_{k} - g_{k}^\mathrm{eq} \right), \label{momentum_flux}
\end{eqnarray}
in the leading order. 
More details of filtered collision procedure can be found in previous studies. \cite{chen2006recovery, zhang2006efficient, latt2006simulating, shan2006kinetic, chen2017lattice, chen2014recovery, otomo2016studies, otomo2018multi}

\section{Validation}
\label{validation}
A set of benchmark tests is conducted using the numerical models proposed in Section~\ref{sec_formulation}.
Throughout this section, $\rho_\mathrm{ratio}=1000$, $M=0.166$, $W=4.0$, and all quantities are in the lattice unit unless otherwise stated.

\subsection{A static droplet simulation for checking the Laplace law and the Galilean invariance}
\begin{figure}
\includegraphics[width=0.9\linewidth]{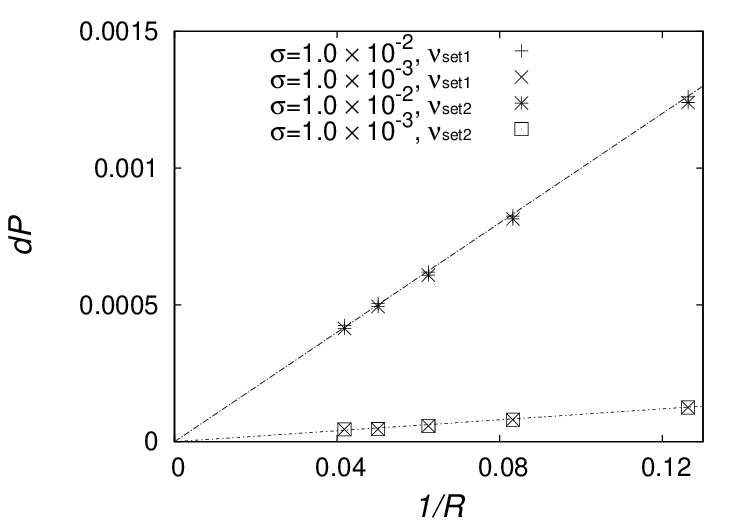}
\caption{ Pressure difference across the interface, $dP$, vs. inverted droplet radius, $1/R$, compared to solid lines following Laplace's law when $\sigma=1.0\times 10^{-2}$ and $\sigma=1.0\times 10^{-3}$. Two sets of viscosities are applied, $\nu_\mathrm{set1} = \left\{ \nu_\mathrm{air}, \nu_\mathrm{water} \right\}= \left\{ 1.67\times10^{-1}, 1.10\times10^{-2}  \right\}$, $\nu_\mathrm{set2} = \left\{ \nu_\mathrm{air}, \nu_\mathrm{water} \right\}= \left\{ 5.56\times10^{-3}, 3.67\times10^{-4}  \right\}$. }
\label{fig_laplaw}
\end{figure}

To check the consistency with Laplace's law and Galilean invariance, a droplet in free space is simulated in a static and in a moving frame.
%
A two-dimensional static droplet of variable initial radius, $R= \left\{ 8,
12, 16, 20, 24 \right\}$, is placed in the center of a domain, whose size is five times
of $R$ and periodic boundaries are assigned to each pair of edges of the domain.
Two sets of viscosities, $\nu_\mathrm{set1}=\left\{ \nu_\mathrm{air}, \nu_\mathrm{water} \right\}= \left\{ 1.67\times10^{-1}, 1.10\times10^{-2} \right\}$  and  $\nu_\mathrm{set2}=\left\{ 5.56\times10^{-3}, 3.67\times10^{-4}  \right\}$ are applied. 
The input surface tension is $\sigma=1.0\times10^{-2}$ or $\sigma=1.0\times10^{-3}$ .
%
In Fig.~\ref{fig_laplaw}, the relationship between $1/R$ and the pressure difference across the interface, $dP$, is shown and compared with lines of slope of $1.0\times 10^{-2}$ or $1.0\times 10^{-3}$ from the Laplace law.
The results match Laplace's law very well and output the consistent $\sigma$ with the inputs even with different viscosity setups.

\begin{figure}
\includegraphics[width=0.9\linewidth]{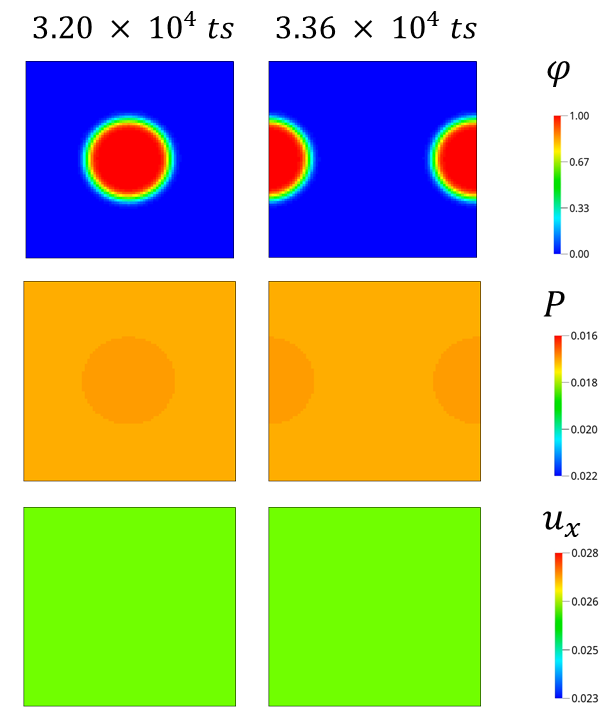}
\caption{ Profiles of $\varphi$ (top), $P$ (middle), and $u_x$ (bottom)  at 3.20$\times 10^4$  (left) and 3.36$\times 10^4$ (right) timesteps using the LB model in Sec.~\ref{sec_formulation} where the initial homogeneous velocity in the $x$-direction of 0.025 is applied. }
\label{fig_dynamicdronew}
\end{figure}

Next, as done in Section~\ref{Intro} with a droplet of $R=16$, the velocity in $x$-direction, $u_x= 0.025$, is initially applied everywhere to simulate in the moving frame and check the consistence with the Galilean invariance. 
All settings are the same as the test in Section~\ref{Intro}.
In Fig.~\ref{fig_dynamicdronew}, the profiles of $\varphi$, $P$, and $u_x$ at 3.20$\times 10^4$  and 3.36$\times 10^4$  timesteps are present.
Unlike Fig.~\ref{fig_dynamicdro}, which uses the $P \mbox{-} \rho u$ scheme, the drop shape remains circular.
Also, the $P$ and $u_x$ profiles seem to be much closer to the expected results of the static frame.
As a result, the LB models in Section~\ref{sec_formulation} exhibit much better Galilean invariance compared to the  $P \mbox{-} \rho u$ scheme. 

\subsection{Poiseuille flow}

\begin{figure}
\includegraphics[width=0.9\linewidth]{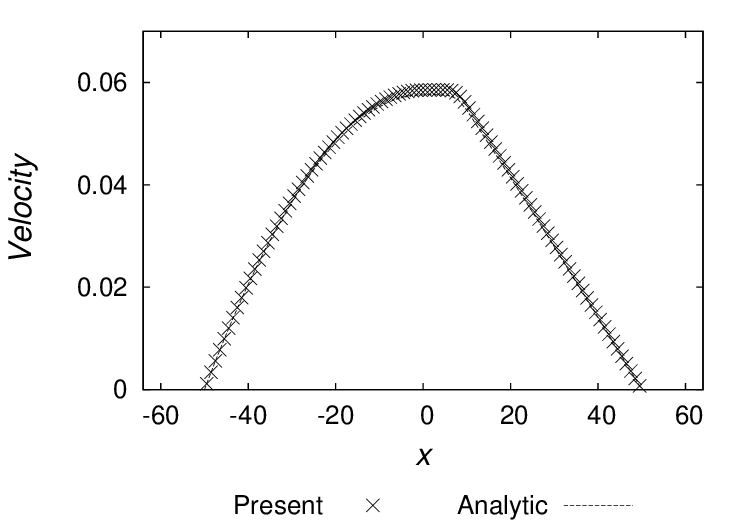}
\caption{ The velocity profile of the two-layer Poiseuille flow using the LB model introduced in Section~\ref{sec_formulation}.  The analytical solution is represented by a line.  }
\label{fig_poiseuille_presen}
\end{figure}

The two-layer multiphase Poiseuille flow discussed in Section~\ref{Intro} is simulated with the LB model of Section~\ref{sec_formulation}. 
All settings are the same as the case in Section~\ref{Intro}.
The simulated velocity profile is compared with the analytical solution in Fig.~\ref{fig_poiseuille_presen}. 
The LB model used here does not have the correction terms for the high orders unlike the $P \mbox{-} \rho u$ scheme in Fig.~\ref{fig_poiseuille_P_rhou}.
Nevertheless, the simulated results agree very well with the analytical solution because the present model is free of the truncation error terms that contaminate the shear stress around the interface as discussed in Section~\ref{Intro}.

\subsection{Head-on droplet collision}
\label{head_on_dro_col}

\begin{figure}
\includegraphics[width=0.9\linewidth]{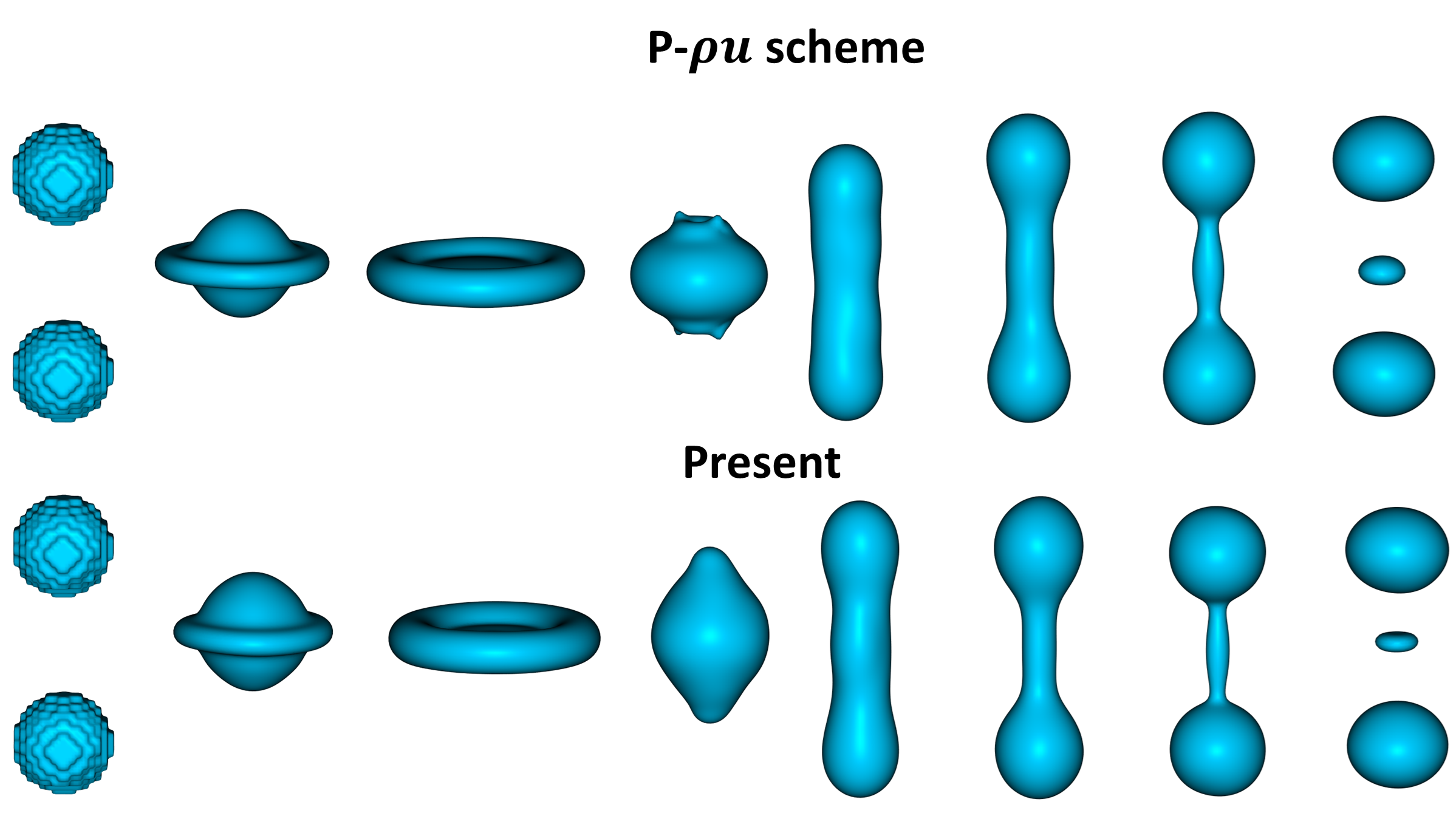}
\caption{ The deformation patterns of the droplets in their head-on collision at $Re =1715$ and $We= 58.4$ using  the  $P \mbox{-} \rho u$ scheme in a previous study \cite{otomo2019improved}(top) and the present model (bottom). The water droplets are described by the iso-surface of $\varphi \ge 0.5$.}
\label{fig_head_on_droplet_col}
\end{figure}

\begin{figure}
\includegraphics[width=0.9\linewidth]{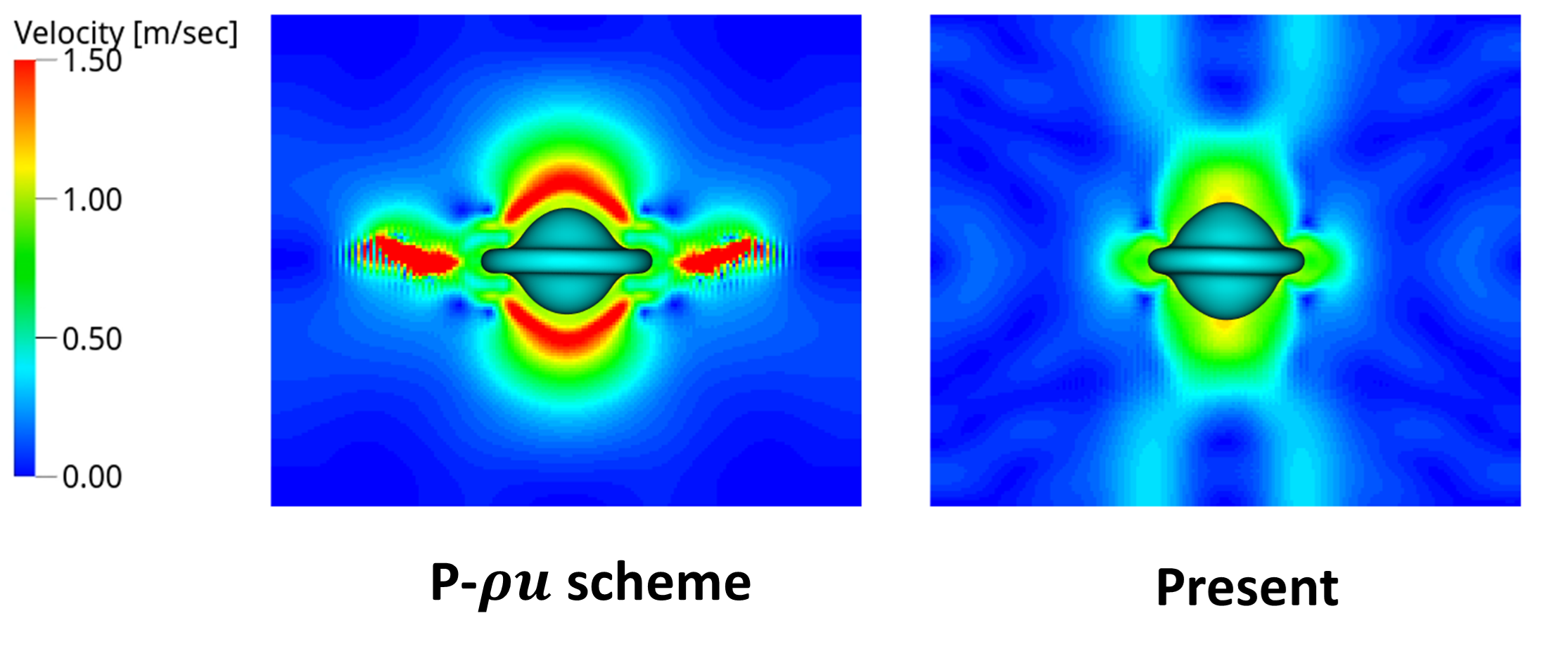}
\caption{Air velocity profiles at a moment of the head-on collision using  the  $P \mbox{-} \rho u$ scheme in a previous study \cite{otomo2019improved} (left) and the present model (right). The water droplet is described by the iso-surface of $\varphi \ge 0.5$.}
\label{fig_head_on_droplet_col_vel}
\end{figure}

The head-on binary droplet collision is simulated and results are compared with the experiments \cite{PhysRevE.80.036301} and with the  $P \mbox{-} \rho u$ scheme in the previous study. \cite{otomo2019improved}
The initial relative velocities of two droplets are set as $2.45 {\;} \mathrm{m}/\mathrm{sec}$.
The droplets' radii are set to $0.35 {\;} \mathrm{mm}$.
The surface tension is $0.072 {\;} \mathrm{N}/\mathrm{m}$ and the viscosities of air and water are $1.51 \times 10^{-5}  {\;} \mathrm{m}^2 /\mathrm{sec}$ and $1.00 \times 10^{-6} {\;} \mathrm{m}^2 / \mathrm{sec}$.
The densities of air and water are $1.0 {\;} \mathrm{kg}/ \mathrm{m}^3$ and $1000 {\;} \mathrm{kg}/ \mathrm{m}^3$.
Hence the Reynold number, $Re$, is $1715$ and the Weber number, $We$, is $58.4$.
Under these conditions, the droplet shows the characteristic behavior due to surface tension and water viscosity. 
%
The simulated results, the temporal sequence of the droplet motion, using the  $P \mbox{-} \rho u$ scheme and the present scheme in Section~\ref{sec_formulation} are shown in Fig.~\ref{fig_head_on_droplet_col}.
After collapsing, the droplets show the disk pattern. Surface tension then causes them to collapse horizontally and the droplet expands vertically. Finally, a droplet splits into two with a satellite droplet in the middle.
Such behaviors observed in the experiment, as shown in Fig.~2 in the paper,\cite{PhysRevE.80.036301} are consistently obtained with both schemes.
The results may indicate that both schemes can reasonably capture the effects of surface tension and water viscosity.
However, upon closer inspection, the air velocity induced by the motion of the droplets is very different between the two schemes, as shown in the snapshots taken at the same moment in Fig.~\ref{fig_head_on_droplet_col_vel}. 
The $P \mbox{-} \rho u$ scheme has greater effects of the moving interface on the airflow field. 
It is possibly because the truncation error terms associated with the density gradient contaminate the momentum transfer across the interface. 
Since the moving water is likely to have the large momentum from the air's point of view, the momentum transfer from water to air must be accurate to avoid artificially accelerating the air.

%

\subsection{Two-dimensional bubble rising}
The two-dimensional bubble rising in a liquid column has been simulated as a canonical benchmark test case with many multiphase flow solvers. \cite{Bubblerising_2009}
The bubble moves upward due to the buoyancy force, and its shape is mainly determined by the balance between the inertial force and the surface tension force.
Here the scheme proposed in Section~\ref{sec_formulation} and the $P \mbox{-} \rho u$ scheme in a previous study \cite{otomo2019improved} are applied and the results are compared.

The bubble rising is simulated at two conditions of $We= \left\{ 10, 125 \right\}$. 
In both cases, a droplet with radius $R=50$ is initially located at $x= 100, y=100$ in the domain of $[ 200 \times 400 ]$. 
Gravity is applied in the negative $y$-direction with an amount of the acceleration $g=1.36 \times 10^{-8}$.
The density of the second component, the bubble, is set to $\rho_2=0.22$.
The other settings are listed in Table~\ref{tab_bubblerise} for two cases where the subscript denotes the component number.   Here the Reynolds number $Re$ and Weber number $We$ are estimated as $Re= \rho_1 U_g 2 R/ \mu_1$ and $\rho_1 U_g^2 2 R/ \sigma$, respectively, where $U_g =\sqrt{2Rg}$.

\begin{table}[]
\caption{Setups for the rising bubble case, where $\rho_1$ is the density of the surrounding heavier fluid, $\mu$ is the dynamic viscosity, and $\sigma$ is the surface tension.}
\label{tab_bubblerise}
\resizebox{1.0\columnwidth}{!}{%
\begin{tabular}{c  cccccc}
\hline
                        & $\rho_1$   & $\mu_1$ & $\mu_2$ & $\sigma$  & $Re$ & $We$ \\ \hline
case1                     & 2.2                          & $7.3 \times 10^{-3}$ & $7.3 \times 10^{-4}$  & $3.0 \times 10^{-5}$   & 35    & 10            \\
case2                     & 220                           &  $7.3 \times 10^{-1}$  & $7.3 \times 10^{-3}$  & $2.4 \times 10^{-4}$   & 35   & 125                                  
 \\ \hline
\end{tabular}%
}
\end{table}

\begin{figure}
\includegraphics[width=0.9\linewidth]{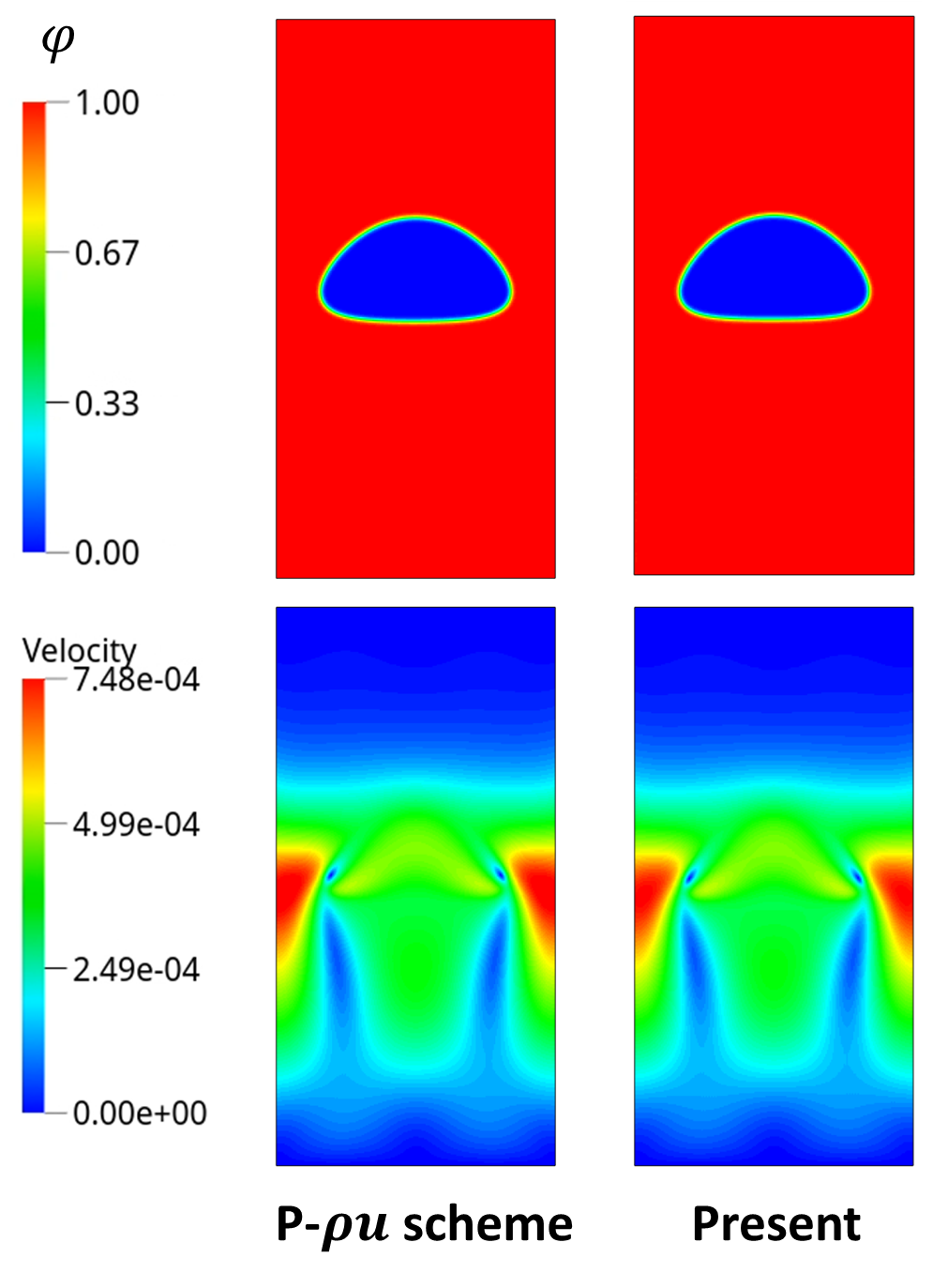}
\caption{ Snapshots of $\varphi$ (top) and velocity magnitude (bottom) profiles at $We=10$ using $P \mbox{-} \rho u$ scheme \cite{otomo2019improved} (left) and the scheme proposed in Section~\ref{sec_formulation} (right).}
\label{fig_bubblerising_lowWe}
\end{figure}

\begin{figure}
\includegraphics[width=0.9\linewidth]{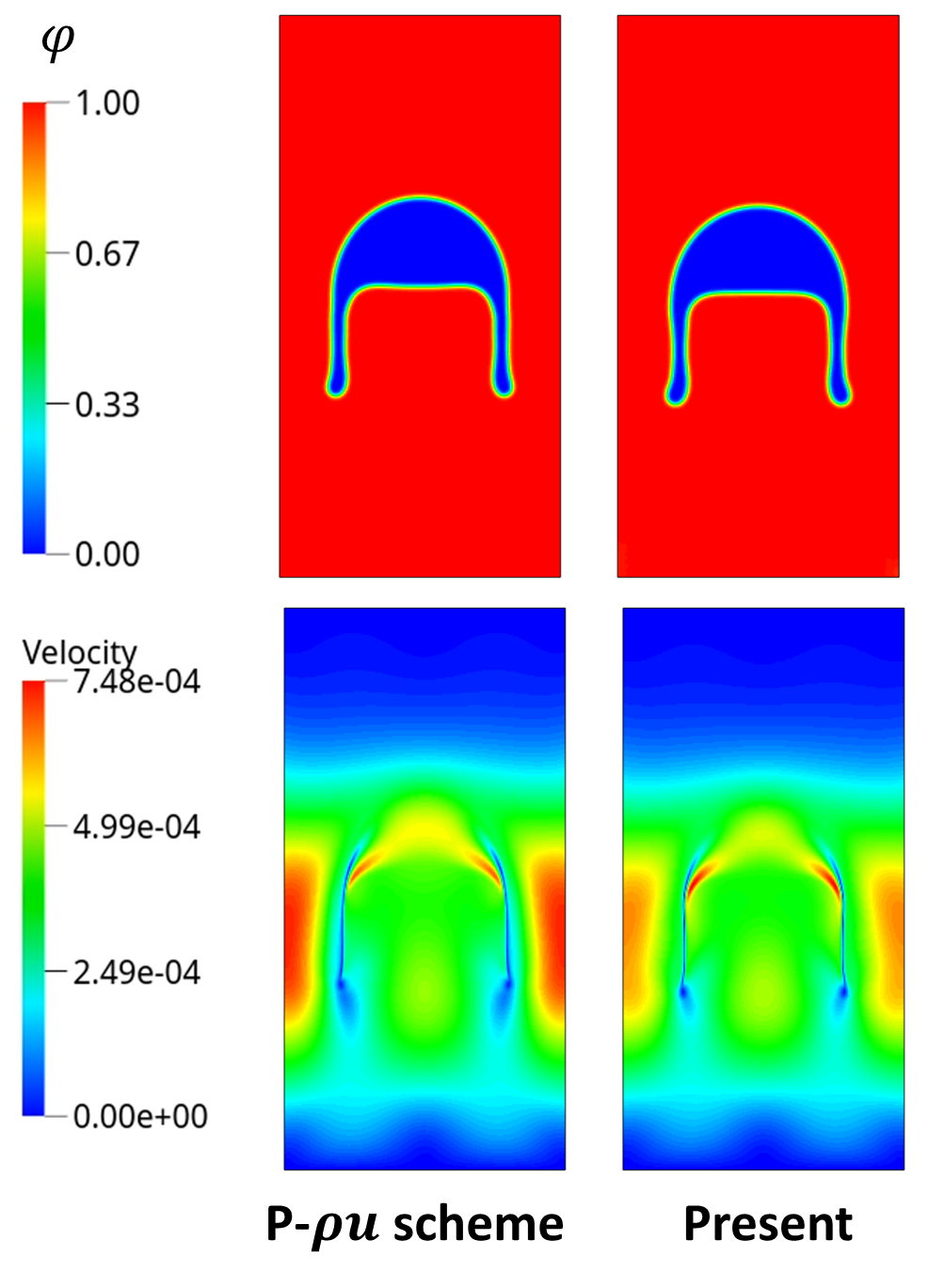}
\caption{ Snapshots of $\varphi$ (top) and velocity magnitude (bottom) profiles at $We=125$ using $P \mbox{-} \rho u$ scheme \cite{otomo2019improved} (left) and the scheme proposed in Section~\ref{sec_formulation} (right).}
\label{fig_bubblerising_highWe}
\end{figure}

The simulated results with the $P \mbox{-} \rho u$ scheme \cite{otomo2019improved} and the scheme in Section~\ref{sec_formulation}  are shown in Fig~\ref{fig_bubblerising_lowWe} and Fig.~\ref{fig_bubblerising_highWe} for the cases of $We=10$ and $We=125$, respectively.
While the bubble is slightly deformed at low $We$, it is largely deformed and has tails in the sides at higher $We$.
These behaviors are consistently captured with these two schemes, as is done with many of the multiphase CFD solvers. \cite{Bubblerising_2009}
We observe small difference between two schemes in the velocity fields as shown in the bottom figures in Fig~\ref{fig_bubblerising_lowWe} and Fig.~\ref{fig_bubblerising_highWe} in contrast to the results in Section~\ref{head_on_dro_col}.
This is probably because the motion of the air, the second component, mainly drives the water, the first component, although the water motion drives the air in the case of Section~\ref{head_on_dro_col}.

\subsection{Three-dimensional dam breaking with block obstacle}
\begin{figure}
\centering
\includegraphics[scale=0.4]{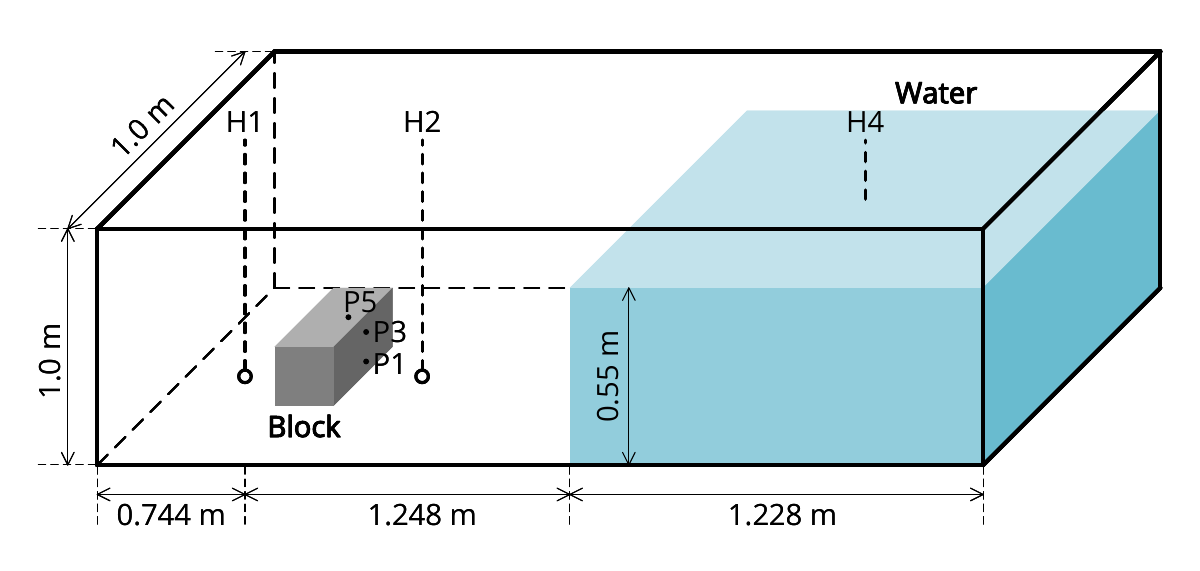}
\caption{The settings of the three-dimensional dam breaking case. The pressure field sensors are placed on the surfaces of the block and are named $P_1$, $P_3$, and $P_5$. The water level sensors are placed at the locations H1, H2, and H4.}
\label{fig:case-dam-break}
\end{figure}

The dam-breaking waves hitting the block and container walls are simulated and compared with experimental results from the Maritime Research Institute Netherlands (MARIN). \cite{kleefsman2005volume}
In a container with size of $ 3.22 {\;}  \mathrm{m} {\times}1 {\;}  \mathrm{m} {\times}1 {\;}  \mathrm{m}$, the water column whose height $H$ is $0.55 {\;} {\mathrm{m}} $ are initially set in the corner of the container.
A block  whose size is $0.161 {\;}  \mathrm{m} {\times} 0.403 {\;}  \mathrm{m} {\times} 0.161 {\;}  \mathrm{m}$  is set in the middle of the downstream side and $0.744 {\;} \mathrm{m}$ from the left edge as shown in Fig.~\ref{fig:case-dam-break}.
On the block, pressure sensors named $P_1$, $P_3$, and $P_5$ are placed at heights of $0.021 {\;} $ m, $0.101 {\;}$ m, and $0.161 {\;}$ m, respectively. 
The sensors of $P_1$ and $P_3$ are placed on the vertical surface and $P_5$ is placed on the horizontal surface at the location $0.021 {\;}$m and $0.101 {\;}$m from the front corner.
The water height sensors named H1, H2, and H4 are placed at a distance of $0.496 {\;} \mathrm{m}$, $0.992 {\;} \mathrm{m}$, and $2.638 {\;} \mathrm{m}$ from the left edge of the container as shown in Fig.~\ref{fig:case-dam-break}.
All sensors are centered in the depth direction.

The water and air densities are set to $ {\rho}_\mathrm{water} = 999.10 {\;} {\mathrm{kg/{m^3}}} $ and $ {\rho}_\mathrm{air} = 1.225 {\;} {\mathrm{kg/{m^3}}} $. 
The kinematic viscosity of water and air is set to ${\nu}_\mathrm{water} = 1.139 {\times}10^{-6} {\;} {\mathrm{{m^2}/{sec}}}$ and ${\nu}_\mathrm{air} = 1.461 {\times}10^{-5} {\;} {\mathrm{{m^2}/{sec}}} $.
The surface tension is ${\sigma} = 0.072 {\;} {\mathrm{N/m}}$.
The resolution is set so that 1 fluid cell resolves $ 5.5 {\;} {\mathrm{mm}}$.
The characteristic velocity, calculated by $\sqrt{2 g H}$ where $g$ is the gravitational acceleration, is set to $0.0289$ in the lattice unit.
All surfaces except the top surface in the domain are set to no-slip walls. And the top surface is set as a pressure fixed boundary at atmospheric pressure.

\begin{figure}
\includegraphics[width=0.9\linewidth]{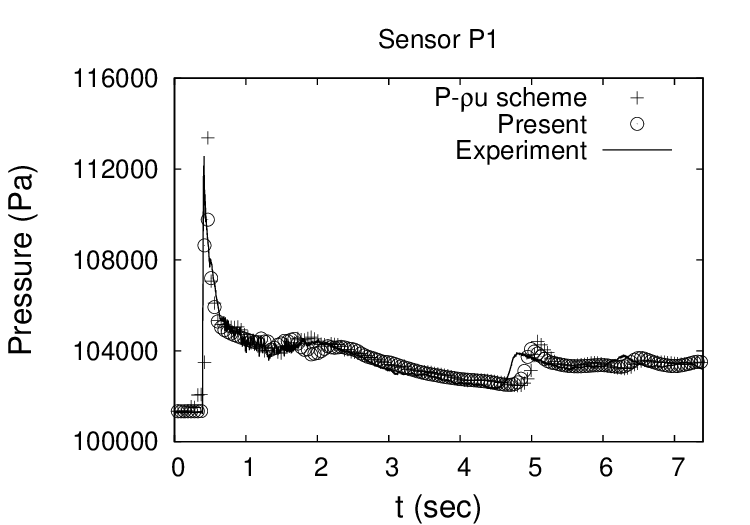}
\caption{ Pressure time history at sensor P1 simulated with the $P \mbox{-} \rho u$ scheme  \cite{otomo2019improved} and the present scheme in Section~\ref{sec_formulation} along with the experimental results\cite{kleefsman2005volume} of the solid line.}
\label{fig_P1sensor}
\end{figure}
\begin{figure}
\includegraphics[width=0.9\linewidth]{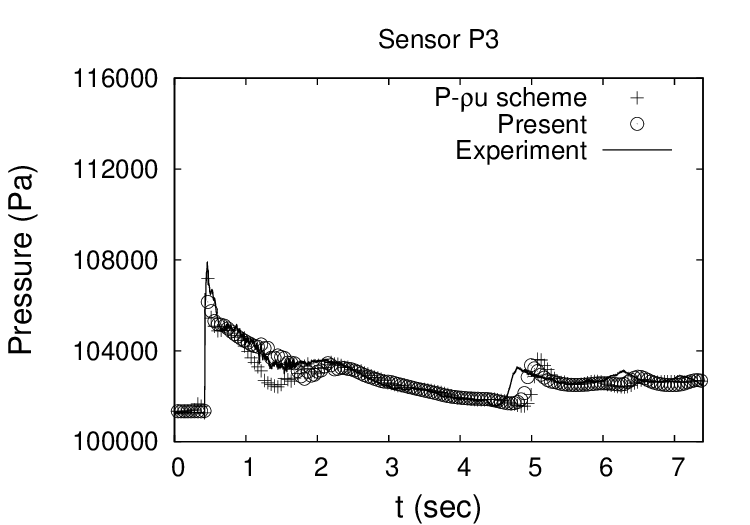}
\caption{Pressure time history at sensor P3 simulated with the $P \mbox{-} \rho u$ scheme  \cite{otomo2019improved} and the present scheme in Section~\ref{sec_formulation} along with the experimental results\cite{kleefsman2005volume} of the solid line.}
\label{fig_P3sensor}
\end{figure}
\begin{figure}
\includegraphics[width=0.9\linewidth]{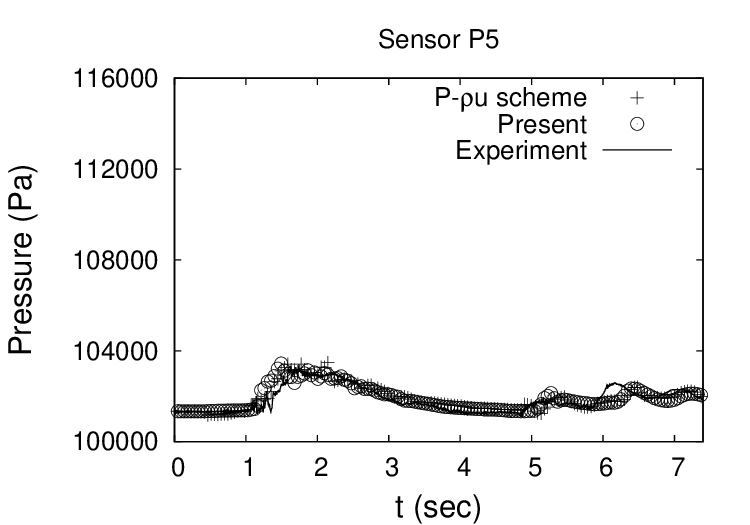}
\caption{Pressure time history at sensor P5 simulated with the $P \mbox{-} \rho u$ scheme  \cite{otomo2019improved} and the present scheme in Section~\ref{sec_formulation} along with the experimental results\cite{kleefsman2005volume} of the solid line.}
\label{fig_P5sensor}
\end{figure}
\begin{figure}
\includegraphics[width=0.9\linewidth]{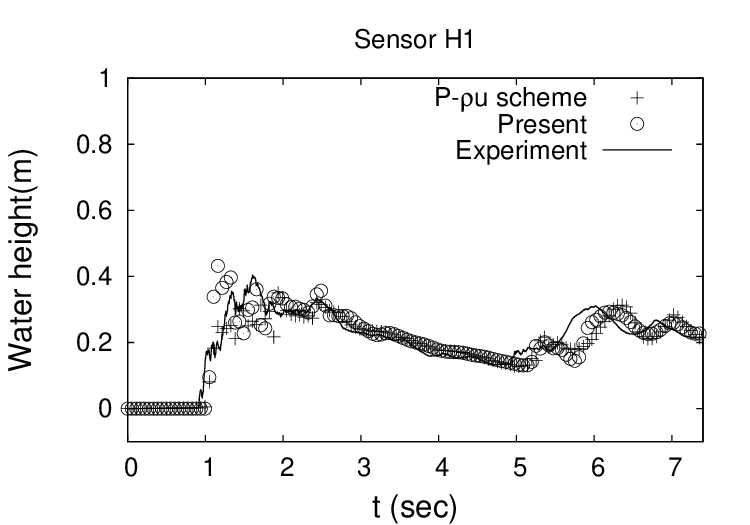}
\caption{Water height time history at sensor H1 simulated with the $P \mbox{-} \rho u$ scheme  \cite{otomo2019improved} and the present scheme in Section~\ref{sec_formulation} along with the experimental results\cite{kleefsman2005volume} of the solid line.}
\label{fig_H1sensor}
\end{figure}
\begin{figure}
\includegraphics[width=0.9\linewidth]{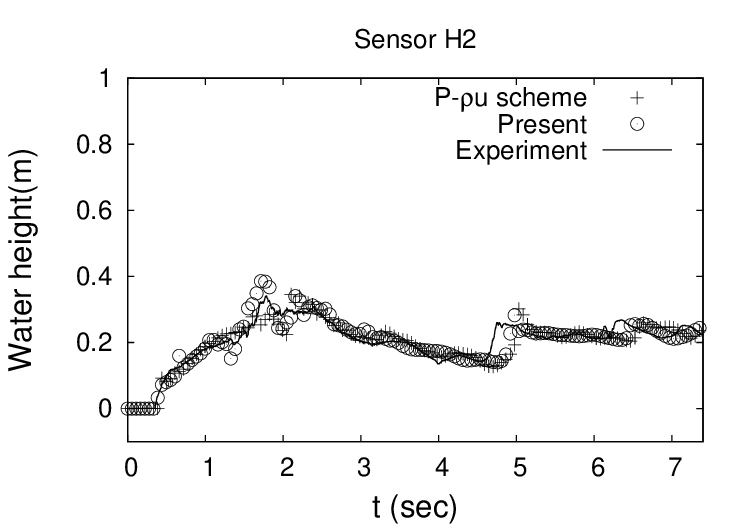}
\caption{Water height time history at sensor H2 simulated with the $P \mbox{-} \rho u$ scheme  \cite{otomo2019improved} and the present scheme in Section~\ref{sec_formulation} along with the experimental results\cite{kleefsman2005volume} of the solid line.}
\label{fig_H2sensor}
\end{figure}
\begin{figure}
\includegraphics[width=0.9\linewidth]{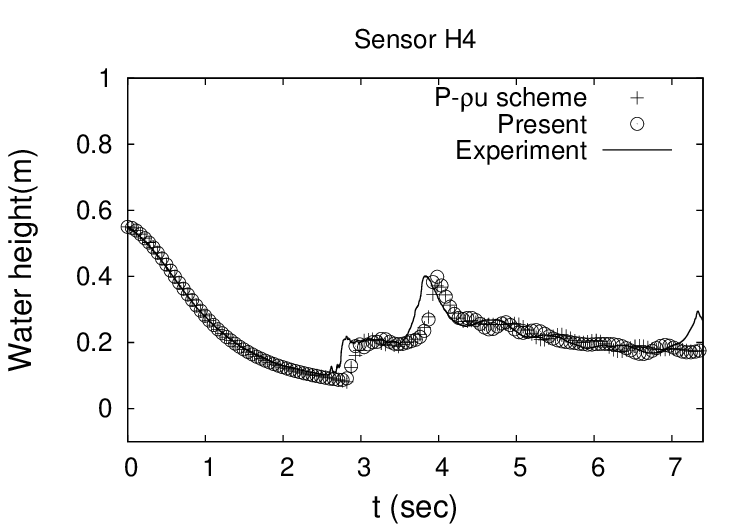}
\caption{Water height time history at sensor H4 simulated with the $P \mbox{-} \rho u$ scheme  \cite{otomo2019improved} and the present scheme in Section~\ref{sec_formulation} along with the experimental results\cite{kleefsman2005volume} of the solid line.}
\label{fig_H4sensor}
\end{figure}
\begin{figure*}
\includegraphics[width=1\linewidth]{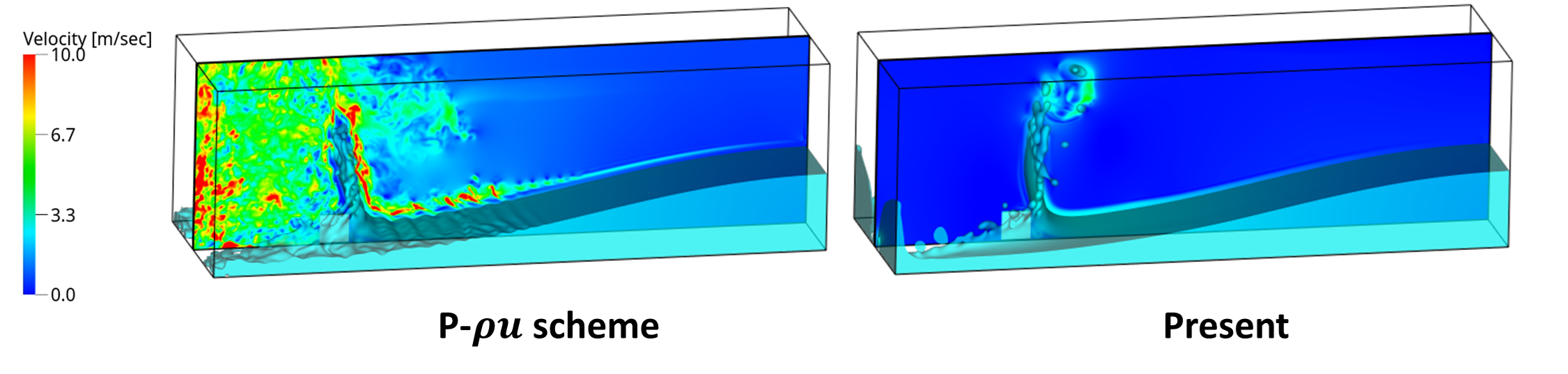}
\caption{Simulated velocity profiles on the central plane at t=0.7 sec and iso-surfaces of $\varphi > 0.5$ using  the $P \mbox{-} \rho u$ scheme\cite{otomo2019improved} and the present scheme  in Section~\ref{sec_formulation}.}
\label{fig_dambreakingwvmf_vel07}
\end{figure*}

The pressure time histories at sensors P1, P3, and P5 with the $P \mbox{-} \rho u$ scheme  \cite{otomo2019improved} and the present scheme  in Section~\ref{sec_formulation} are compared with the experimental results in Fig.~\ref{fig_P1sensor}, Fig.~\ref{fig_P3sensor}, and Fig.~\ref{fig_P5sensor}.
Similary, the water height at sensors H1, H2, and H4 with the $P \mbox{-} \rho u$ scheme  \cite{otomo2019improved} and the present scheme  in Section~\ref{sec_formulation} are compared with the experimental results in Fig.~\ref{fig_H1sensor}, Fig.~\ref{fig_H2sensor}, and Fig.~\ref{fig_H4sensor}.
Both of the $P \mbox{-} \rho u$ scheme \cite{li2023examining} and the present scheme show reasonable agreement with the experiment.
For example, the first peak due to the first attack on the block around $t=0.5$ sec is accurately captured with the correct timing.
Some of the delays of the secondary waves around $t=4.5$ sec can be seen in these pressure histories.
These delays of the water motion can also be seen in the water height histories at H2 around $t=4.5$ sec and at H4 around $t=4.0$ sec.
As discussed in the study, \cite{li2023examining} there may be some extra dissipation through the water dynamic motion with this level of spatial resolution as the other solvers. \cite{green2017sloshing,marsooli20143}
The main difference between the $P \mbox{-} \rho u$ scheme \cite{otomo2019improved} and the present scheme can be seen in the flow field of the air region.
In Fig.~\ref{fig_dambreakingwvmf_vel07}, the velocity contours at 0.7 sec in the centre plane are shown with the iso-surface of $\varphi > 0.5$.
It is a moment after the water first hits the block when a large amount of the kinetic energy of the bulk water is converted to potential energy.
As can be clearly seen, the air is greatly accelerated by the movement of the water. 
Considering the characteristic speed of water $\sqrt{2 g H}=3.28 {\;} \mathrm{m}/\mathrm{sec}$, the speed with $P \mbox{-} \rho u$ scheme is at an extremely high level. Here we use the wide range of colored contours, such as one from $0 \mathrm{m}/\mathrm{sec}$ to $10 \mathrm{m}/\mathrm{sec}$, to show this acceleration.
The tip splash pattern has also changed.
This may be due to the differences in momentum transfer from water to air resulting from the truncation error terms discussed in Section.~\ref{Intro}. 

\section{Summary}
\label{summary}
The lattice Boltzmann based formulation for solving the hydrodynamic equations of pseudo-incompressible multiphase flows with high density ratios is discussed.
Due to the high density ratio, the high-order truncation error terms related to the spatial density gradient can be significant.
For example, as shown in a previous study, even the fourth derivative order terms can cause the obvious accuracy problem around the interface \cite{otomo2019improved}.
To control such high-order terms by adding the correction terms is challenging because higher-order terms are more complicated.
The $P/\rho \mbox{-} u$ scheme proposed in a study \cite{fakhari2017improved} can improve such issues related to the velocity field.
However, since such truncation error terms still exist in the pressure equation, this scheme may have the absolute pressure value dependence that is unacceptable for the incompressible flow framework.
To overcome this problem, we develop the $P \mbox{-} u$ scheme by mapping the lattice kinetic scheme (LKS) based models \cite{INAMURO201655,INAMUROBOOK} to the generic LB models.
After some simplification and implementation of the filter collision operator, reasonable stability and accuracy are achieved, as shown in many validation cases in Section~\ref{validation}.
Indeed, in the static and dynamic droplet cases, we confirmed that Galilean invariance and the absolute pressure dependence are significantly improved.
One of the most important practical improvements over the $P- \rho u$ scheme is the accuracy of the air flow field induced by the water motion, as shown in the droplet collision and dam-breaking wave cases in Section~\ref{validation}. 
This may be due to improvements in the momentum transfer between two phases by avoiding fatal truncation error terms in the hydrodynamic equations.
\edit{In this study, the LB formulation for the hydrodynamic equations has been focused while using an LB model based on one of the popular phase field equations along with certain model parameter sets. In the future, a more detailed study of the coupling with solvers for different phase field equations and the parameter study are desired. }
 
\section*{Acknowledgements}
\edit{We thank the organization of the "33rd Discrete Simulation of Fluid Dynamics (DSFD)" of the "ETH Zurich" in "Zurich, Switzerland" held over "9 - 12 July 2024" for creating the platform on which and for bringing together the audience to which this work was first presented. H.O. would like to thank Prof. Taehun Lee and Dr. Daniel Lycett-Brown for helpful discussions.}

\nocite{*}
\bibliography{manuscript}

\end{document}